\documentstyle[12pt]{article}
\def\cad{{\cal D}}
\def\bA{{\bar A}}
\def\d{\partial}
\def\bd{{\bar\partial}}
\def\a{\alpha}
\def\th{\theta}
\def\g{\gamma}
\def\G{\Gamma}

\def\N{\nabla}
\def\bN{\bar\nabla}
\def\tb{{\bar\theta}}

\def\de{\delta}
\def\P{\phi}
\def\f{\varphi}

\def\h{\chi}
\def\bh{{\bar\chi}}

\def\l{\lambda}

\def\L{\Lambda}
\def\bL{{\bar\Lambda}}
\def\s{\sigma}
\def\p{\psi}

\def\is{\equiv}
\def\bD{{\bar D}}
\def\ka{K\"ahler\ }
\def\kaa{K\"ahler}

\def\V{{e^V}}
\def\iv{{e^{-V}}}
\def\q{{\ ,\ \ \ }}
\def\qqq{\qquad\qquad\ }

\def\pmb#1{\setbox0=\hbox{#1}%
\kern.0em\copy0\kern-\wd0
\kern-.04em\copy0\kern-\wd0
\kern.08em\copy0\kern-\wd0
\kern-.04em\raise.0433em\box0 } 	

\def\half{{\textstyle{1 \over 2}}}
\def\ihalf{{\textstyle{i \over 2}}}

\def\bop#1{\setbox0=\hbox{$#1M$}\mkern1.5mu
        \vbox{\hrule height0pt depth.04\ht0
        \hbox{\vrule width.04\ht0 height.9\ht0 \kern.9\ht0
        \vrule width.04\ht0}\hrule height.04\ht0}\mkern1.5mu}

\begin{document}

\newcommand{\inv}[1]{{#1}^{-1}} 

\renewcommand{\theequation}{\thesection.\arabic{equation}}
\newcommand{\beq}{\begin{equation}}
\newcommand{\eeq}[1]{\label{#1}\end{equation}}
\newcommand{\ber}{\begin{eqnarray}}
\newcommand{\eer}[1]{\label{#1}\end{eqnarray}}
\begin{center}
        August, 1996		\hfill    ITP-SB-96-39\\
			        \hfill    USITP-96-09\\
				\hfill    hep-th/9608112\\

\vskip .1in

{\large \bf Yang-Mills fields for Cosets}
\vskip .2in

{\bf Johan Grundberg} \footnotemark \\

\footnotetext{e-mail address: ejg@vana.physto.se}

{\em Department of Mathematics and Physics \\
 M\"alardalens H\"ogskola \\
Box 833 \\
S-721 23 V\"aster{\aa}s SWEDEN} \\

\vskip .15in

{\bf Ulf Lindstr\"om} \footnotemark \\

\footnotetext{e-mail address: ul@vana.physto.se}

{\em  Institute of Theoretical Physics \\
University of Stockholm \\
Box 6730 \\
S-113 85 Stockholm SWEDEN}\\

\vskip .15in

{\bf Martin Ro\v cek} \footnotemark \\

\footnotetext{e-mail address: rocek@insti.physics.sunysb.edu}

{\em Institute for Theoretical Physics \\
State University of New York at Stony Brook \\
Stony Brook, NY 11794-3840 USA}\\
\vskip .1in
\end{center}
\vskip .4in
\begin{center} {\bf ABSTRACT } \end{center}
\begin{quotation}\noindent
We consider theories with degenerate kinetic terms such as those that
arise at strong coupling in $N=2$ super Yang-Mills theory. We compute
the components of generalized $N=1,2$ supersymmetric sigma model
actions in two dimensions. The target space coordinates may be 
matter and/or Yang-Mills superfield strengths.
\end{quotation}
\vfill
\eject
\def\baselinestretch{1.2}
\baselineskip 16 pt
\noindent

\section{Introduction}
\setcounter{equation}{0}

Yang-Mills theory
\cite{YM} is the essential ingredient in our understanding of all
fundamental interactions.  Part of its beauty comes from its rigid
geometric structure: Once a gauge group is chosen, the basic degrees of
freedom of the gauge sector are specified.  In this note, we present a
generalization of this structure, and find Yang-Mills fields that in some
sense are gauge fields for cosets.

It has long been known that in gauge theories with scalar fields,
nonminimal couplings may modify the gauge-field kinetic terms.  In
particular, positive definite kinetic terms were constructed for
noncompact groups in \cite{jl,hklr}.  Here we consider another
possibility: semi-definite terms which give rise to gauge theories where
part of the gauge group is auxiliary. As an example, consider an
$SU(2)$ gauge theory with an adjoint representation scalar field $\f^i$.
Then, using the methods of \cite{hklr}, the most general
gauge invariant action with no more than two derivatives is
\ber
S_0&=&\int d^Dx \left[ g_1(|\f |)\left(\delta_{ij}-\frac{\f^i\f^j}{|\f
|^2}\right) (f_{MN}^i f_{MN}^j+h_1(|\f
|)\nabla_M\f^i\nabla_M\f^j)\right. \cr
&&\left.\qquad +g_2(|\f |) \left(
\frac{\f^i\f^j}{|\f |^2}\right) (f_{MN}^i f_{MN}^j+h_2(|\f
|)\nabla_M\f^i\nabla_M\f^j)+V(|\f |) \right]\ ,
\eer{n0}
Minimal coupling implies $g_1=g_2=h_1=h_2=1$, and the usual
positive definite case arises when these functions are all positive.
We want to consider the case when $g_1$ or $g_2$ vanishes; in the first
case, only the gauge field for a $U(1)$ subgroup remains dynamical, and
the $SU(2)/U(1)$ gauge fields become auxiliary; this is analogous to the
discussion of ``composite gauge-fields'' (see, {\em e.g.,} \cite{com}).
When $g_2$ vanishes, we have a new situation: the $U(1)$ gauge-field is
auxiliary, and only the gauge-fields for the coset $SU(2)/U(1)$ remain
dynamical.

In \cite{hklr}, a general theory of invariant nonminimal kinetic terms was
developed; the main object there was to construct positive definite kinetic
terms for noncompact gauge groups.  However, exactly the same techniques can
be used to construct degenerate kinetic terms such as those described above.
Indeed, using the techniques of \cite{hklr}, we can couple gauge fields to
nonlinear sigma-models in arbitrary dimensions, and select which gauge fields
are dynamical and which are auxiliary.  As in the examples above, we can
``keep'' either the fields of a subgroup, or of a coset.  In this sense, we
have constructed a gauge theory for cosets. A possible application of
this would be to effective actions with fewer gauge fields than local
gauge symmetries. 

Though in two dimensions an action such as (\ref{n0}) is renormalizable,
in higher dimensions it is more interesting when considered as the low
energy limit of an effective action.  Indeed, specific functions
$g_1,g_2,h_1,h_2$ were given for the $D=4, N=2$ super Yang-Mills low energy
effective action in \cite{sw}.  As shown by two of us, though $g_2$ is
always positive and hence the $U(1)$ gauge field remains dynamical for
all values of $<|\f |>$, along a certain curve, $g_1$ vanishes, and then
actually changes sign; we interpret this as a signal that the coset gauge
fields become nondynamical and disappear from the spectrum
\cite{lr}.

Extended super Yang-Mills theories contain physical scalars as
superpartners of the gauge-fields; this means that in superspace, there
are scalar super field-strengths which can be used to construct
nonlinear sigma models. This has been studied for $D=4$, $N=2$ theories (the
``K\"ahlerian vector multiplet'' \cite{kvm,hklr}), where it gives
rise to a limit of ``special \ka geometry''.  Here we consider the
analogous sigma-models in two dimensions (similar constructions
can be done in higher dimensions). For $N=1$, $D=2$, the super field-strength
is a scalar superfield that transforms in the adjoint representation of
the gauge group, and can be used to coordinatize {\em any\/} manifold
that is invariant under the gauge-symmetry and has dimension less than
or equal to the dimension of the group. Similarly, for $N=2$, $D=2$, the super
field-strength is a complex twisted (covariantly) chiral superfield
(again in the adjoint representation) \cite{ghr},
and can be used to coordinatize {\em any \kaa\/} manifold that is invariant
under the gauge-symmetry and has dimension less than or equal to
twice the dimension of the group. (Generalizations exist for
actions that include twisted super Yang-Mills multiplets; these
give rise to manifolds with torsion).

In the next section we collect the basics of $N=1$ Yang-Mills theory
in $D=2$ and give the reduction to $N=0$ of a general nonlinear action
including matter-fields. In Sec.\ 3.\ we turn to $N=2$, recapitulate the
basic definitions and representations, and reduce a general action,
containing both covariantly chiral and twisted chiral fields, to
$N=1$. We find that it takes on the same form as for the usual
nonlinear $\sigma$-model. In both sections, we give $SU(N)$ examples
analogous to (\ref{n0}).

\section{$N=1$}
\setcounter{equation}{0}

We first briefly review $N=1$, $D=2$ super Yang-Mills theory. The covariant
superspace derivatives $\nabla_\pm\equiv D_\pm+i\Gamma_\pm$ satisfy the
algebra
\beq
(\nabla_\pm )^2 = \nabla_{\pm\pm}\q\{\nabla_+,\nabla_-\}=F\is F^iT_i\ ,
\eeq{nis1}
where $\nabla_{\pm\pm}\equiv\d_{\pm\pm}+i\Gamma_{\pm\pm}$, $\d_{++}\equiv
\d$, $\d_{--}\equiv\bd$ are the usual (anti)holomorphic
derivatives on the world-sheet and $T_i$ are the Lie algebra generators
satisfying $\left[T_i,T_j\right]=c_{ij}^{\ \, k}T_k$. The super
field-strength $F$ is an unconstrained scalar except that the Bianchi
identities imply:
\beq
\nabla_+\nabla_-F=-[\nabla_{++},\nabla_{--}]\ .
\eeq{bian}
The standard component fields are defined by
$$
\f\is F |\q A\is\Gamma_{++}|\q\bA\is\Gamma_{--}|\q
\l_+\is\nabla_+F |\q\l_-\is\nabla_-F | \ ,
$$
\beq
f\is\d \bA-\bd A +i[A,\bA ]= -\nabla_+\nabla_-F |\ .
\eeq{n1c2}

We consider a superspace action for the Yang-Mills multiplet for a gauge
group $G$
\beq
S_{N=1}=\frac1{2\pi}\int d^2z \nabla^2
\left[\left(g_{ij}(F)+b_{ij}(F)\right)
\nabla_+F^i\nabla_-F^j\right]\ ,
\eeq{n1act}
where $i,j$ are group indices, $g$ is some $G$-invariant metric, and
$b$ is a two-form that is invariant modulo exact terms. The minimal
case occurs when $b=0$ and $g$ is the killing metric $k_{ij}$.
Typical {\em non}minimal terms are:
\beq
g_{ij}(F)= g_0(F) k_{ij} + g_1(F) F^iF^j + ...
\eeq{gnon}
\beq
b_{ij}= b_0(F)F^k c_{ijk} + ...
\eeq{bnon}

We may also consider nonminimal couplings to extra matter
multiplets. These are described by scalar superfields $\P^\mu$ that
coordinatize some manifold which admits an action of the gauge
group generated by Killing vector fields $k^\mu_i(\P)$.
The scalars have component expansions:
\beq
X^\mu\is\P^\mu |\q \psi_\pm^\mu\is\nabla_\pm\P^\mu |\q S^\mu\is
\half [\nabla_+,\nabla_-]\P^\mu |\ ,
\eeq{n1cfi}
where the gauge covariant derivatives are
\beq
\nabla\P^\mu \is D\P^\mu +i\Gamma^ik^\mu_i\ .
\eeq{n1covdir}
The nonminimal kinetic term for the Yang-Mills multiplet
may depend on these scalars, and the general formalism of \cite{hklr}
can be applied.  Furthermore, the gauge fields couple to the scalar
kinetic term as described in \cite{hs}; in many cases, this is
simply a matter of minimal coupling using the covariant derivatives
(\ref{n1covdir}) in an action analogous to $S_{N=1}$ (\ref{n1act}) with
$\{F^i\}\to\{F^i,\P^\mu\}$.

Using the definitions of the components (\ref{n1c2}), we find that the action
$S_{N=1}$ (\ref{n1act}), including matter-fields $\P^\mu$, has the
component expansion 
\ber
S_{N=1}&=&{1\over{2\pi}}\int d^2z\
\left\{
(g_{\a \beta}(Y)+b_{\a\beta}(Y))(-\N_{++}Y^{\a}\N_{--}Y^{\beta})\right.
\cr&&\cr
&&\qqq+g_{\a\beta}(\p_+^\a \cad_{--}\p_+^\beta+
\p_-^\a\cad_{++}\p_-^\g)
\cr&&\cr
&&\qqq+\half R_{(-)\a\beta\g\delta}\p_+^\delta \p_-^\beta
\p_+^\g \p_-^\a-i\p_+^{\a}\l_-^i \tilde k_{i\alpha}+i\l_+^i
\tilde k_{i\a}\p_-^{\a}
\cr&&\cr
&&\qqq\left.-\left( S_\a+\ihalf b_{\beta\a}\f^i k_i^\beta -
\G_{+\beta\g}^{\ \ \ \de} g_{\de\a}\p_+^\g\p_-^\beta\right)^2\right.
\cr&&\cr
&&\qqq\left.-\frac14\left( 
\f^i\tilde k_{i\a}\right)^2+i\f^i\left(\tilde
k_{i\a}\right)_{;\beta}\p_+^\beta\p_-^\a\right\}
\cr&&\cr
&&\hbox{\hfill}
\eer{n1comp}
where the symbolic squaring of expressions within parenthesis is
shorthand for the scalar product in the metric $g_{\a \beta}$
 and
\ber
\cad_{\pm\pm}\p_\mp^\beta &\is&
\N_{\pm\pm}\p_\mp^\beta + (\N_{\pm\pm}Y^\de)\G_{(\pm)\g\de}^{\qquad\beta}
\p_\mp^\g\ ,\cr&&\cr
\N\p^\beta &\is&  \partial\p^\beta
+i\G^ik^\beta_i,_\a\p^\a\ , \cr&&\cr
\tilde k_{i\a} &\is& (g_{\a\beta}+b_{\a\beta})k_i^\beta \ .
\eer{cader}
The connections are defined as
\beq
\G_{(\pm) \a \beta}^{\qquad\g} \is \G_{(0) \a \beta}^{\qquad\g} \pm T_{\a
\beta}^{\quad\g}
\eeq{condef}
with $\G_{(0) \a \beta}^{\qquad\g} $ the metric (Christoffel) connection
and the torsion
defined as
\beq
T_{\a\beta\g}\is \half (b_{\a\beta ,\g}+b_{\g [\a ,\beta ]}).
\eeq{torsion}

The curvature tensor formed from $\G_{(\pm) \a \beta}^{\qquad\g}$ is
denoted by $
R_{(\pm)\a\beta\mu\nu}$,  $;$
denotes the $\G^-$-covariant derivative w.r.t. $Y$,
and we have used a collective notation for the fields:
\beq
Y^\a\is \left(\begin{array}{c}
\f^i\cr
X^\mu\cr
\end{array}
\right),
\quad
\p^\a_\pm\is \left(\begin{array}{c}
\lambda^i_\pm\cr
\p^\mu_\pm\cr
\end{array}
\right),
\quad
S^\a \is \left(\begin{array}{c}
f^i\cr
S^\mu\cr
\end{array}
\right),
\quad
k_i^\a\is
\left(\begin{array}{c}
c_{\ ik}^j\f^k\cr
k_i^\mu\cr
\end{array}
\right),
\eeq{collfield}
where $c_{ik}^j$ are the structure constants of the Lie algebra under
consideration.
To separate the $N=0$ Yang Mills field strength $f^i$ from the
auxiliary matter field $S^\mu$ we rewrite the $S^\alpha$ term in
(\ref{n1comp}) as follows
\ber
-\left[S^\mu + i\half b_\a{}^{\mu} \f^ik_i^\a - \G_{+\a\beta}^{\quad \mu}
\p_+^\beta\p_-^\a
+g^{\mu\a}g_{\a i}\left(f^i+i\half b_\beta {}^{i}\f^jk_j^\beta
-\G_{+\beta\gamma}^{\quad i}\p_+^\gamma\p_-^\beta\right)\right]^2&&\cr&&
\cr
-\left(f^i+i\half b_\beta {}^{i}\f^jk_j^\beta
- \G_{+\beta\gamma}^{\quad i}\p_+^\gamma\p_-^\beta\right)
\tilde g_{ik}
\left(f^k+i\half b_\beta {}^{k}\f^jk_j^\beta
- \G_{+\beta\gamma}^{\quad k}\p_+^\gamma\p_-^\beta\right),&&\cr
&&
\eer{square}
where we have used the metric $g_{\mu\nu}$ in the first square
and  $\tilde g_{ik} \is (g_{ik}-g_{i\mu}g^{\mu\nu}g_{\nu k})$.

We are particularly interested in cases when the metric $\tilde g$ is
degenerate; then, though the action is invariant under the full gauge
group, some of the component gauge and scalar fields are auxiliary. We
consider two examples for the gauge group $SU(N)$: Let 
\ber
g_{ij}=\left( {1\over {|F|^2}}\right)\left[ g_1
\left(\delta^{ij}-{{F^iF^j}\over
{|F|^2}}\right)+g_2{{F^iF^j}\over {|F|^2}} \right]\ ,
\eer{gsu2}
where $g_1=0, g_2=1$ or $g_1=1, g_2=0$. We include no matter
fields and put $b_{ij}=0$. The bosonic part of the lagrangian in 
(\ref{n1comp}) is then proportional to (\ref{n0}) with $h_1=h_2=1$ and
$g_1, g_2$ as above. (The proportionality factor is $1/{|F|^2}$.)

\section{$N=2$}
\setcounter{equation}{0}

The same phenomenon as described in the previous section occurs
for $N=2$, $D=2$ super Yang-Mills.  In part because this is how we
discovered the phenomenon, and in part because of the curious interplay
of these ideas and \ka geometry, we now give a detailed discussion.

In two dimensions, the $N=2$ Yang-Mills supermultiplet is described by
a
gauge-covariantly twisted chiral superfield \cite{ghr}.  It is well
known that ordinary chiral superfields can be interpreted as the complex
coordinates of a \ka manifold \cite{Z}.  In particular, the superspace
Lagrangian
\beq
K=\ln (\Phi^\mu\bar\Phi^\mu)\q \mu = 1,...,n+1,
\eeq{K}
gives rise to a $\s$-model on the manifold $CP(n)$, with $\Phi^\mu$
the natural homogeneous coordinates (chiral superfields).  Because of \kaa 
-invariance, the
component action depends not on all the $\Phi^\mu$, but only on their
ratios. If we replace
$\Phi^\mu$ by the field strength of a nonabelian gauge multiplet, we find a
system that does not depend on all the usual gauge
fields, but rather only on certain combinations; nevertheless, we
maintain the full gauge invariance of the system.

We begin by reviewing some background material.

\subsection{Chiral, twisted chiral, and variant superfields}

Extended supersymmetric multiplets are generally described by constrained
superfields (see, for example, \cite{book}).  In our case ($D=N=2$), we
will work with chiral superfields $\Phi$ that satisfy the constraints
\beq
\bD_{\pm}\Phi=0\q D_{\pm}\bar\Phi=0\ ,
\eeq{fi}
where the spinor derivatives $D,\bD$ as usual satisfy the supersymmetry
algebra
\beq
\{ D_+,\bD_+\} =\d\q \{ D_-,\bD_-\} =\bd\ .
\eeq{alg}
The constraints (\ref{fi}) have the general solution:
\beq
\Phi=i\bD_+\bD_-\Psi\q \bar\Phi=iD_+D_-\bar\Psi\ .
\eeq{ccsol}
The chiral superfield reduces to a complex unconstrained $N=1$ scalar
superfield.

We will also work with twisted chiral superfields $\h$ that satisfy
twisted constraints \cite{ghr}
\beq
\bD_+ \h = D_- \h = 0\q D_+ \bh= \bD_- \bh = 0\ ,
\eeq{chi}
which can be solved by
\beq
\h=i\bD_+D_-\Psi\q \bh=iD_+\bD_-\bar\Psi\ .
\eeq{tcsol}
Just as for the chiral case, when we project to $N=1$ superspace, the
twisted chiral superfield reduces to a complex unconstrained $N=1$
scalar superfield.

As explained in \cite{var}, though these
solutions are general, they are not unique: there exist variant multiplets
satisfying the same constraints.  For example, a variant multiplet is
found by constraining $\Psi=\bar\Psi\equiv V$.  The effect of this is to
replace a complex auxiliary field by a real field and the
divergence(curl) of a vector field. Since, in two dimensions, vector
fields are not dynamical, except for possible topological issues, this
does not change the physical content of the theory.  In particular, for
the twisted case, the variant multiplet has a familiar interpretation
\cite{ghr}: It is the superfield strength of a $U(1)$ gauge multiplet.

\subsection{$D=2$, $N=2$ Yang-Mills theory}

In the $N=2$ case, the gauge covariant derivatives obey the
constraints:
\ber
\N_\pm\is D_\pm +i\G_\pm\q &\N_{\pm\pm}=\{\N_\pm ,\bN_\pm\}&\q
\N_{\pm\pm}\is\d_{\pm\pm}+i\G_{\pm\pm}\ ,\cr&&\cr
\{\bN_+,\N_-\} ={\cal F}\q &\{\N_+,\bN_-\} =\bar{\cal F}&
\q\N_\pm^2=\bN_\pm^2=0\ ,
\eer{Field}
where ${\cal F}$ is a complex
superfield strength. The Bianchi identities imply that ${\cal F}$ is a gauge
covariantly {\em twisted chiral\/} superfield ({\em c.f.\/} \ref{chi}):
\beq
\bN_+{\cal F}=\N_-{\cal F}=0\ .
\eeq{gctc}
A convenient solution to the constraints can be found in a chiral
representation, {\it i.e.\/}, in a representation where the gauge
parameter is a chiral superfield $\L$:
\beq
\bN_\pm=\bD_\pm\q \N_\pm=\iv D_\pm \V\ ,
\eeq{eV}
where $V$ is an unconstrained prepotential that transforms as
\beq
\V '=e^{i\bL}\V e^{-i\L}\ .
\eeq{trans}
Note that in chiral representation, since the gauge parameter is chiral,
and hence complex, $\N\ne (\bN)^{\dag}$, etc. 

We can also consider a twisted super Yang-Mills theory with
a twisted chiral gauge parameter which couples to covariantly twisted
chiral matter and has a covariantly chiral field strength. The twisted
theory is constructed by interchanging $\N_-$ with $\bar \N_-$ in
(\ref{Field}-\ref{eV}) above.

The $N=1$ components of the untwisted multiplet
are defined as:
\beq
F^i\is {\cal F}^i+\bar {\cal F}^i, \quad \phi^i\is -i( {\cal F}^i-\bar
{\cal F}^i),
\eeq{comp1}
where $F^i$ is the $N=1$ Yang-Mills field strength and $\phi^i$ is
$N=1$ matter. 

\subsection{$CP(n)$ $\s$-models}

Sigma-models in $N=2$ superspace are described by a super-Lagrangian
that is just a function of chiral (and/or twisted
chiral) superfields and their complex conjugates: $K(\Phi ,\bar\Phi )$. In
the case with only chiral or twisted chiral superfields, the superfields
can be interpreted as complex coordinates on a \ka manifold with \ka
potential $K$ \cite{Z}. The metric is just the complex Hessian of $K$,
and hence is unchanged if one adds any holomorphic function to it; in
superspace, this \kaa -invariance arises because any function of purely
chiral superfields is itself chiral, and hence annihilated by the full
superspace measure $D^2\bD^2$.

A particularly simple description exists of the $\s$-model on $CP(n)$;
the superspace Lagrangian is:
\beq K=\ln (\Phi^\mu\bar\Phi^\mu)\q \mu= 1,...,n+1,
\eeq{K2}
where $\Phi^\mu$ are homogeneous coordinates.  Because of
\kaa-invariance, the component action depends not on all the $\Phi^\mu$,
but only on their ratios; equivalently, the model is invariant under
the gauge transformation
\beq
\Phi^\mu\to \L\Phi^\mu\q \bar\Phi^\mu\to \bL\bar\Phi^\mu\ ,
\eeq{ktrans}
for an arbitrary chiral superfield $\L$. Then one can choose the gauge
$\Phi^{n+1}=1$.

We now consider generalized $\s$-models constructed out of $N=2$
superfields.

\subsection{\ka manifolds and gauge theories}

In this section we derive
the $N=1$ super-component action for a general superspace
Lagrangian
\beq
L_{N=2}=K(\Phi^A,\bar\Phi^B)\ ,
\eeq{KFF}
where $A\in (\mu,i)$, and $\Phi^\mu$ ($\Phi^i$) are covariantly
(twisted) chiral superfields; $\Phi^\mu$ is thus either a matter field
covariant with respect to an untwisted Yang-Mills symmetry or a field
strength of a twisted Yang-Mills multiplet whereas $\Phi^i$ is either
a matter field covariant with respect to a twisted Yang-Mills symmetry
or a field strength of an untwisted Yang-Mills multiplet.
In deriving the $N=1$ action we will make the
assumption that $K(\Phi^A,\bar\Phi^B)$ is invariant under local gauge
transformations. This makes it possible to use covariant superspace
derivatives in the superspace measure. In the language of \ka geometry
we are assuming that $K$ is invariant under the isometry we have
choosen to gauge. This is not the most general possibility, but is
sufficient for our purposes.

To rewrite the invariant Lagrangian (\ref{KFF}) in $N=1$ language, we
define new covariant derivatives
\beq
\N^1_\pm\is \bN_\pm + \N_\pm ,\quad \N^2_\pm\is i(\bN_\pm - \N_\pm).
\eeq{nabla1}
In terms of these derivatives the covariant (twisted) chirality
constraints become
\beq
\N^2_{\pm}\Phi^A=-i\varepsilon_\pm (A)\N^1\Phi^A, \quad
\N^2_{\pm}\bar\Phi^A=i\varepsilon_\pm (A)\N^1\bar\Phi^A,
\eeq{covtwi}
where no summation over $A$ is intended and the sign-factor is
\beq
\varepsilon_\pm(\mu)=1, \quad \varepsilon_\pm(i)=\pm 1.
\eeq{sign}
We take the $N=1$ algebra to be the one spanned by $\N^1$ (c.f. (\ref{nis1})):
\beq
\{\N^1_\pm ,\N^1_\pm\}=2\N_{\pm \pm}, \quad \{\N^1_\pm ,\N^1_\mp\}=F.
\eeq{n1alg}
Using (\ref{nabla1}), the action corresponding to (\ref{KFF}) may be written
\ber
S&=&\int d^2zd^2\th d^2\tb K(\Phi^A,\bar\Phi^B) = \int
d^2z\N_{+}\N_{-}\bN_{+}\bN_{-}K(\Phi^A,\bar\Phi^B)|\cr
&=&-{1\over4}\int d^2z
\N^1_{+}\N^1_{-}\N^2_{+}\N^2_{-}K(\Phi^A,\bar\Phi^B)|,
\eer{n1int}
 where
$|$ denotes the projection onto the $\th_2$-independent part.
Letting the $\N^2$'s act on $K$ we find, using (\ref{covtwi}),
\beq
S=-\half\int d^2zd^2\th\left( G_{IJ}
+B_{IJ}\right)\N_{+}\P^I\N_{-}\P^J,
\eeq{GB}
where $\N\is \N^1$ and $\th\is \th_1$ is the corresponding Fermi
coordinate. The new indices have the range $I\in (A,\bar A)$ and
the target space metric and antisymmetric tensor field are
\beq
G_{IJ}=\left(\begin{array}{cccc}0&K_{\mu\bar \nu}&0&0\cr
K_{\bar \mu\nu}&0&0&0\cr
0&0&0&-K_{i\bar j}\cr
0&0&-K_{\bar ij}&0
\end{array}
\right)
\eeq{G}
\smallskip
\beq
B_{IJ}=\left(\begin{array}{cccc}0&0&0&-K_{\mu\bar j}\cr
0&0&-K_{\bar \mu j}&0\cr
0&K_{i\bar \nu}&0&0\cr
K_{\bar i\nu}&0&0&0
\end{array}
\right).
\eeq{B}
\smallskip
The $N=0$ component action is found by substituting 
(\ref{GB},\ref{G},\ref{B}) into (\ref{n1comp}); 
a form of it has been previously given in \cite{ggw}.

Above we have not interpreted the twisted covariantly chiral fields as
either matter-fields or Yang-Mills fields. If $\Phi^A$-fields are
all matter fields, the $N=1$ component fields are simply their
$\th_2$-independent parts. If, for some $A$, $\Phi^A={\cal F}^i$ (the
$N=2$ field strengths), we also have to display the $N=1$ components
(\ref{comp1}).  For the special case with no $\Phi^\mu$'s and all
$\Phi^i={\cal F}^i$, and the \ka potential $K=ln\Phi^i\bar\Phi^i$ we
obtain $B_{IJ}=0$ and a $G_{IJ}$ that gives

\ber
S&=&\int d^2zd^2\th\left\{\left({1\over{(F^2+\phi^2)}}\right)
\left[\left(\de ^{ij} -
{{F^iF^j+\phi^i\phi^j}\over{(F^2+\phi^2)}}
\right)\left(\N_{+}F^i\N_{-}F^j+\N_{+}\phi^i\N_{-}\phi^j\right)\right.
\right.\cr
&&\cr
&&\left.\left.\qquad\qquad\qquad\qquad\qquad
-{{F^{[i}\phi^{j]}}\over{(F^2+\phi^2)}}
\N_{[+}F^j\N_{-]}\phi^i\right]\right\}.
\eer{lnex}

\vskip .3in \noindent
{\bf Acknowledgements} \vskip .2in \noindent
We are grateful to J.Gates, M.Grisaru and R. von Unge for comments on the 
manuscript. EJG would like to thank the ITP
at Stony Brook for its hospitality. MR would like to thank the ITP at the
University of
Stockholm for its hospitality. The work of MR was supported in part by
NSF grant No.
PHY 9309888. The work of UL was supported in part by NFR grant
No. F-AA/FU 04038-312 and by NorFA grant No 966003-0.


\end{document}